\documentclass[%
 reprint,
 amsmath,amssymb,
 aps,
]{revtex4-2}

\usepackage{graphicx}
\usepackage{dcolumn}
\usepackage{bm}
\usepackage{gensymb}
\usepackage{amsmath}
\usepackage{caption}

\begin{document}

\preprint{APS/123-QED}

\title{Memristive behavior of functionalized graphene quantum dot\\ and polyaniline nanocomposites }

\author{D.P. Pattnaik}
\email{d.pattnaik@lboro.ac.uk}
\author{A.T. Bregazzi}%
\author{P. Borisov}%
\author{S.Savel'ev}%
\email{s.saveliev@lboro.ac.uk}
\affiliation{%
 Physics Department, Loughborough University, Loughborough, LE11 3TU, United Kingdom\\
}%


\author{A.B. Siddique}
\author{M. Ray}%
\email{mallar.ray@tec.mx}
\affiliation{
 School of Engineering and Sciences, Tecnológico de Monterrey, Monterrey, Nuevo Leon, Mexico\\
}%

\author{N. Banerjee}
\email{n.banerjee@imperial.ac.uk}
\affiliation{%
 Department of Physics, Blackett Laboratory, Imperial College London, London, SW7 2AZ, United Kingdom\\
}%


\date{\today}

\begin{abstract}
Zero-dimensional graphene quantum dots (GQD) dispersed in conducting polymer matrix display a striking range of optical, mechanical, and thermoelectric properties which can be utilized to design next-generation sensors and low-cost thermoelectric. This exotic electrical property in GQDs is achieved by exploiting the concentration of the GQDs and by tailoring the functionalization of the GQDs. However, despite extensive investigation, the nonlinear resistivity behavior leading to memristive like characteristic has not been explored much. Here, we report electrical characterisation of nitrogen functionalized GQD (NGQD) embedded in a polyaniline (PANI) matrix. We observe a strong dependence of the resistance on current and voltage history, the magnitude of which depends on the NGQD concentration and temperature. We explain this memristive property using a phenomenological model of the alignment of PANI rods with a corresponding charge accumulation arising from the NGQD on its surface. The NGQD-PANI system is unique in its ability to matrix offers a unique pathway to design neuromorphic logic and synaptic architectures with crucial advantages over existing systems

\end{abstract}

\maketitle


\section{\label{sec:level1}Introduction}

Rapid progress in logic and memory devices in the last two decades has enabled processing and storage of large volumes of data. This has led to a several breakthroughs in computing including the recent rise of AI driven by the success of deep learning. However, there are distinct bottlenecks in the current silicon-based microelectronics which critically limits further progress: for example, complementary metal-oxide semiconductor (C-MOS) technology is challenging to scale down beyond the sub-10 nm regime and logic processing and memory being separately located – the so-called von Neumann bottleneck,\cite{Burks1982} limits the energy efficiency and speed .\cite{Meijer2008,Bondyopadhyay1998,Frank2001} These challenges require a radical rethinking of the current computing mechanisms, and any proposed alternate devices should demonstrate increased functional complexity as well as improved interconnectivity beyond the von-Neumann architectures. 
Neuromorphic computing is one such promising avenue which is modelled after that human brain and amongst the devices which can make it a reality are memristors. Since its discovery, memristors \cite{DiVentra2009,Pickett2013} have been widely recognized as an emerging device which are scalable below 10 nm, with tuneable resistance states, ease of fabrication and integration to current semiconductor technology leading to a promising boost to future neuromorphic technologies.\cite{Adhikari2012,Jo2010,Pattnaik2023} 

In memristors various effects lead to the change in resistance when a current or voltage is applied, and these history-dependent changes rely on mechanisms like current/heat driven phase transitions or formation of conductive filaments between the memristor electrodes. \cite{Gabbitas2023,Pattnaik2023,sun2019understanding}. The resistance changes in all memristor devices are driven by either volatile or non-volatile switching events.\cite{zhou2022volatile} For volatile switching, memristor always return to the same state (usually with high resistance) when the voltage is turned off allowing them to be used for artificial neurons when a volatile memristor is placed in parallel with a capacitor and in series to a load resistance .\cite{pattnaik2023temperature,Pattnaik2023,ushakov2021role} Interestingly, these devices generate spiking when driven by slowly changing or DC voltages \cite{Pattnaik2023,midya2019artificial} making them ideal candidates for neuron-like elements which can be used in neuromorphic computing. In contrast to volatile memristors, non-volatile memristors can memorize previous activities even if the driving (either voltage or current) is switched off, thereby offering an opportunity to use them as artificial synapses.\cite{zhou2022volatile}  Combining artificial neurons and synapses offers a path to realize memristor-based artificial neural networks.\cite{Wang2018,yao2020fully}

However, current material systems available for the design of these non-volatile memristors have one major limitation: the possible change of resistance is usually determined by the number of available states that the system can reliably switch to, and availability of multiple states are relatively rare (see, e.g.,Ref. \cite{stathopoulos2017multibit}). This limitation in the number of available states restricts the learning capability of neuromorphic circuits leading to an active search of memristive platforms with continuous variation of resistance. In this context, an attractive class of materials with diverse electrical memory features such as analog and threshold resistive switching are diblock copolymers and conducting polymers. These materials have attracted significant interest for the ability to scale down the devices by overcoming the challenges of fabrication in the lithographic limit.\cite{ouk2003epitaxial,li2000dense,cheng2001formation,arunachalam2018polymer,yusoff2019graphene,demin2015electrochemical,black2007polymer,ito2000pushing,naulleau2019optical}  

Conductivity of these polymers can be easily tuned over a wide range by suitable doping which has the advantage that they are easy to fabricate and have been extensively studied for photovoltaic applications, light emitting diodes, transistors.\cite{kulkarni2004electron,coakley2004conjugated,gunes2007conjugated}  
Here, we report on the experimental memristance behavior in nitrogen functionalized graphene quantum dots (NGQDs) incorporated in a polyaniline (PANI) matrix. Prior studies have indicated the memristive potential of PANI and GQD systems individually, where both PANI and GQD exhibit hysteretic I-V characteristics.\cite{Kalita2013,demin2015electrochemical} The investigation into the electrical properties of graphene quantum dots (GQDs) dispersed within a polyaniline (PANI) matrix, with a specific focus on memristive characteristics, has not yet been undertaken extensively. Hence, we investigate the hysteretic properties of nitrogen-functionalized GQD embedded in polyaniline matrix (NGQD-PANI) which allows tuning the properties of the composite system simply by altering the concentration of the nano constituents. N-functionalized GQDs were preferred over other kinds of GQDs due to the benefits that N-doped carbon nanomaterials provide in terms of charge transfer, storage and interaction, higher electrical conductivity, and better surface hydrophilicity. \cite{Chabot2014,Lin2015} Although the potential of PANI-GQD/NGQD system as candidate memristive material have been suggested earlier,\cite{Luk2014,Siddique2021}  to the best of our knowledge, there is no record of pinched hysteresis, which is widely considered to be the “fingerprint” of a memristive system. \cite{Chua2014} Remarkably, the NGQD-PANI nanocomposites demonstrate pinched hysteresis current voltage characteristic without any threshold switching, thus, offering continuous variation of resistance, needed for hardware-level implementation of artificial networks requiring highly adjustable synapses, resembling even more biological artificial synaptic networks. \cite{Erokhin2007} The magnitude of this memristance behavior strongly depends on the concentration of the NGQD in PANI matrix, thereby allowing the intriguing possibility to fabricate artificial synapses with different weights in resistance change. 
Although non-volatile memory effects in graphene has been studied previously, \cite{Echtermeyer2008,Wang2014} there has been only a few explorations \cite{Dodda2021,Schranghamer2020} towards emulation of graphene as memristive synapses. The usage of functionalized graphene QDs for memristive synapses is a new concept, and in this work we demonstrate embedding nitrogen- functionalized graphene QDs in a polyaniline conducting matrix (NGQD-PANI) prepared using standard hydrothermal method discussed earlier.\cite{Siddique2021,Ferrari2013}

\section{\label{sec:level2} Experimental Section}
\subsection{Materials and Methods}
The NGQD dispersed PANI nanocomposite samples were prepared by pelletizing dried powders of solution processed NGQD-PANI colloids. First, the NGQDs were prepared by hydrothermal treatment of citric acid and ethylene diamine following standard protocol.  The process involved autoclaving (8 hours at $\sim$ 180 $\degree$ C) a mixture of 4.2 g of citric acid and 3.6 g of ethylene diamine in 100 ml of DI water, in an oil bath. In the next step, 1, 5 and 10 weight $\%$ of NGQDs (calculated with respect to aniline monomer) were administered during the polymerization of aniline to form the nanocomposites. Requisite amounts of NGQDs were first mixed with a surfactant, cetyl trimethyl ammonium bromide (CTAB) in aqueous solution of camphor sulfonic acid (CSA, 0.17 M) and stirred for approximately 5 min. Subsequently, 4 g of purified aniline was added, and the stirring was continued for another 1 hour. During the stirring process, 500 mL of ammonium persulfate (APS, 0.085 M) was slowly added dropwise to the mixture. The entire process of in-situ synthesis of NGQD-PANI nanocomposite was carried out in an ice bath and the solution temperature was maintained around 10 0C under continuous magnetic stirring to facilitate slow polymerization. After 24 hours of stirring at low temperature, black precipitates were obtained. The precipitates were filtered and washed with ethanol and DI water until the filtrates turned colorless. The residue was then dried under vacuum for 24 hours at 80 0C to get the NGQD-PANI powder. An aliquot of 200 mg of this powder for each of the NGQD concentrations were pelletized using a hydraulic press by applying 150 kg/cm2 pressure for 10 minutes in a 13 mm diameter die. A control sample made of only PANI, without any NGQD was prepared following the same protocol. All the chemicals and reagents used in the synthesis were of analytical grade procured from Merck, India and Fisher Chemicals.

	Characterization: The microstructure of NGQDs and NGQD-PANI nanocomposites were investigated by a JEOL JEM-2200 FS high resolution transmission electron microscope (HR-TEM) and a JSM-7610F, JEOL field emission scanning electron microscope (FE-SEM), respectively. X-ray diffraction (XRD) studies were carried out by a Brucker D8 advanced diffractometer using Cu K$\alpha$1 radiation. Fourier transform infrared (FTIR) spectra were recorded by JACSO 4700 LE. An Omicron Multiprobe (Omicron NanoTechnology Gmbh, UK) spectrometer fitted with an EA125 (Omicron) hemispherical analyzer was used for carrying out x-ray photoelectron spectroscopy (XPS) measurements. Al-KK$\alpha$ was used as a monochromatic source operated at 150 W along with pass energy analyzer set at 40 eV. Sample neutralization was done by a low-energy electron gun (SL1000, Omicron) fixed at -3 V. 
 
For the current-voltage (I-V) measurements at room temperature, the pellets were cut into 4 × 3 × 1.4 mm3 rectangular sized specimens, which were mounted onto an 8-pin puck (copper pads) and contacted with copper wires using silver paint. For the low temperature measurements, the pellets were cut into 3 × 2 × 1.4 mm3 and mounted on a smaller 8-pin puck and hosted in a liquid N$_2$ filled cryostat. A thermocouple was connected close to the sample to measure the sample temperature. A Keithley 4200-SCS was used to obtain the I-V, where an I-V sweep was obtained between -10V and +10V.

\section{Results and Discussion}
The overall structural features of the NGQDs and the NGQD-PANI nanocomposites are summarized in Figure 1. The TEM micrographs of the NGQDs reveal nearly circular-shaped particles with diameters ranging from 1 nm to 4 nm, an example is shown in Figure 1(a). The crystalline character of the dots is evident from the fringes in the high magnification image shown as inset, Figure 1(a). A fringe width of 0.32 nm corresponds to the (002) interlayer spacing of graphene .\cite{Tachi2019} The quality of the graphitic lattice in the NGQDs was assessed using Raman spectroscopy as shown in Figure 1 (b). The peaks located at around 1410 cm$^{-1}$ and 1530 cm$^{-1}$ are associated with the "D band" and "G band", respectively. The G band corresponds to the in-plane optical phonon modes with sp$^{2}$ hybridization, while the D band represents the phonon modes for defects with sp$^{3}$ hybridization within the lattice. The shift in frequency from its usual position are due to the size reduction.\cite{Dervishi2019}

However, The ID/IG ratio, which represents the integrated intensities of the D and G bands respectively, is a significant indicator of both defect density and crystallite size,\cite{Ferrari2013}, was estimated to be 0.84, which confirms the disordered graphitic structure. The typical graphitic features additionally exhibited by presence of the second order characteristics, which correspond to the 2D, D+G, and 2G combination overtones are marked in Figure 1(b). \cite{Dervishi2019} In addition, one additional band appears at 1410 cm$^{-1}$, which corresponds to the incorporation of N in the graphitic structure, hence confirming the formation of NGQDs. \cite{Kuzhandaivel2021}
The as-synthesized PANI, on the other hand, is dominated by rod-like quasi 1D structure along with some detectable agglomerates, as seen from the FESEM micrograph in Figure 1(c). Interestingly, the PANI rods become more aligned with increasing concentration of NGQDs due to adjunction or interwinding which we have discussed in detail elsewhere.\cite{Luk2014} Figure 1(d) shows this enhanced alignment of PANI rods for the 10 $\%$ NGQD sample which appear to be connected by a network of agglomerated structures. Since the NGQDs are almost two orders of magnitudes smaller than PANI, the individual dots are not resolvable in the SEM micrographs showing the distribution of PANI. In the TEM image shown in Figure 1(e), we can clearly identify the quantum dots dispersed in a PANI backbone.

Further confirmation of this aligning of the PANI rods with increasing NGQD concentration is reflected in the XRD profiles shown in Figure 1f. The XRD patterns are evidently dominated by typical features of repeating benzenoid and quinoid units in PANI. \cite{Dervishi2019} All the XRD peaks consistently become sharper with increase in GQD concentration, suggesting an increase in effective crystallinity due to enhanced alignment. The bonding between the different ligands in NGQDs and in the nanocomposites are revealed in the FTIR spectra shown in Fig.  1(g). The NGQDs are characterized by signature peaks at 1376 and 1226 cm$^{-1}$ due to bending and stretching modes of C-NH, and C-N, respectively. \cite{Ferrari2013,Kuzhandaivel2021} Both these peaks disappear in all the spectra the NGQD-PANI due to bonding between the repeating units of the aniline and NGQDs which in turn leads to the network of quasi aligned PANI as observed in XRD and electron micrographs.\cite{miao2013high}  In all the nanocomposite samples, the typical PANI absorption peaks due to the repeating units of quinoid and benzenoid rings appear at 1572 cm$^{-1}$ and 1492 cm$^{-1}$, respectively.\cite{zhang2014solid} The ratio of intensities of these two peaks is nearly 1:1, indicating mixed oxidation state of PANI. \cite{Qu2014} The other two peaks at 1300 cm$^{-1}$ and 1238   cm$^{-1}$ are due to the stretching vibration modes of C-N secondary aromatic amines and C=N stretching vibrations.\cite{palaniappan2002emulsion,devadas2018effect}  We also note that the relative intensity of polaron absorption peak at 1110 cm$^{-1}$ increases with increasing NGQD concentration indicating enhanced polaron vibration.\cite{bhandari2018polyaniline}

Figure 2 shows the room- temperature I-V characteristics of the pelletized NGQD-PANI composites for different NGQD concentrations. These I-V characteristics are the average of ten I-V cycles. All the I-V curves demonstrate a hysteresis behavior where there is a gradual change in resistance. The reason for this gradual change is further discussed in the next section based on a phenomenological model we propose, but here we provide an explanation for the hysteresis based on the molecular structure of NGQD.
The typical structure of chemically synthesized NGQDs consists of C=C core along with functional groups formed by C, N, O, and H, such as NH2, OH, C-OH, C-H, C-O, etc., as evidenced from the structural investigation discussed above. The surface functional groups introduce energy states, which typically appear between the C=C related core energy levels or bands. Charge transport through such systems is therefore affected by these states, particularly those appearing close to the Fermi energy. Reportedly, these surface-species-related states act as traps for charge carriers where charges are trapped and de-trapped under external bias\cite{Siddique2021,siddique2021excitation}. The hysteretic characteristics observed in NGQDs can be largely attributed to these trap states and their changes created by the functional groups. \cite{Luk2014} Changing the concentration of NGQD in PANI is expected to cause variations in the trap states due to variations in the electron delocalization densities. As a result, the hysteresis behavior due to trapping and de-trapping of charges, is also affected. Increased charge de-trapping and partial dispersal can also be traced back to voltage-induced conformational changes in the structure of the conducting polymer, which leads to the increase in area of the hysteresis loop. \cite{chhabra2020pani} In a PANI-NGQD system, the carrier pathways during charge transport are not only defined by polaronic transport through a typical $\pi$-conjugated system, but also by charge transfer between PANI and NGQD islands. Under an external voltage of suitable magnitude, intramolecular charge transfer moves electrons from PANI to the quasi-continuous bands as well as the surface traps of NGQD. This transfer leads to partial delocalization of electrons in the $\pi$-conjugated system and increases the conductivity leading to lower resistance states. Reversal of the voltage reverts to high energy states but not all trapped charges are released when the voltage is reversed leading to hysteresis.

We utilize this gradual change in resistance as a function of NGQD concentration to selectively tune the resistance depending on the history of the applied voltage. To demonstrate this, voltage to our NGQD devices was applied in the following way. A positive voltage was swept from 0 to +1V. After this, the voltage was cycled to x V, where x= {0.4, 0.5, 0.6, 0.7} and shown in Figure 3 (a-c). We call x as the reset voltage.

This is a unique behavior for the memristor as it opens an additional degree of freedom for the control and tuning of resistance. For memristive devices that demonstrate threshold-based abrupt switching, cycling the voltage below the threshold value and above, results in the same measured resistance, or simply the same I-V characteristic. This is because, in these devices, the resistance change is associated with the formation and rupture of the conduction channel and cycling the voltage below and above only causes partial or complete formation or rupture of the channel. However, in the NGQD-PANI devices, the memristive property originates from the charge trapping in the PANI matrix by the NGQD centers.\cite{Kalita2013} These charge trapping is due to the NGQD nanostructures, which undergo alignment when an external field is applied. As the alignment is governed by various factors such as density of nanorods, the interaction between the nanorods and the matrix, the corresponding resistance shows a behavior that reflects the nanorods alignment in the PANI matrix.

We explain the memristive behavior in the following way: Due to the charge trapping by the NGQD, when an external voltage is applied, most likely the nanorods start to align along the field direction. However, before complete alignment, when the voltage is cycled back, the rods start to realign. When the voltage is cycled up again, the rods start to align from this position and continue to orient. It is observed that for the 5$\%$ NGQD device, the resistance across all the x values showed minimum change when the voltage was cycled. However, for the other higher concentration, a difference in measured current can be seen for different values of x. With different reset values, the re-orientation cycle shows unique resistance value. As the concentration is further increased to 20$\%$, the difference in resistance is more pronounced. This is a clear indicator that the concentration of the nanorods is key towards to selectively tuning the resistance. Furthermore, we also observe that the area of the hysteresis in the I-V characteristic shows a non-monotonic dependence on concentration.

To understand this further, voltage ramps were applied to all the devices. (Figure 3 (d-f)) But in this case the reset voltage (x) was set to 0.2V. This was to ensure that the rods have mostly realigned to their initial state. It is observed that area of the loops are reduced and also the measured current is also reduced as the concentration is increased. 

To study the effect of temperature towards the dry fiction and nanorod alignment, we performed similar voltage sweeps as shown in inset of Fig. 3 (d) for a separate sample with 5$\%$ concentration in a cryostat. Figure 4(a, b) shows the measured IV characteristic for different voltage ramps (series of voltage sweeps (see fig. 3 (d-f) for details)).

 A clear effect of temperature towards the hysteresis is observed. For RT, the measured current for a voltage ramp penetrates through the preceding ramp, with a wide hysteresis loop behavior. While for 77K, the loops are on top of the other, with a clear separation between each voltage ramp. The loops also show a suppressed behavior with a higher measured current compared to the RT case. 
We discuss this temperature dependence with more details in the next modelling section.
\section{Modelling memristive response of NGQD-PANI composites}
To simulate memristor I-V hysteresis we assume that the devices can be modelled as an ensemble of conducting nanorods trapped by an electric potential. The hysteresis and non-linearity in I-V characteristics for NGQD-PANI, are typically attributed to surface- or edge-related states functioning as charge trapping centers \cite{Kalita2013} which give rise to the electric potential.

In our system, the main contribution to the device resistance likely originates from tunnelling of electrons between conducting rods and from/to memristor electrodes. Electric field can push the charged rods (dipoles) to align decreasing tunnelling gaps between them and memristor electrodes, thus decreasing the resistance. Assuming that electrons can tunnel from one electrode to the rod of length l and from the rod to another electrode we can estimate the resistance of the memristor as  $R = \exp(l(1 - \cos \theta))$ ; with the resistance  normalized by minimum resistance when rode is connected the electrodes ($\theta$=0) and rod distance is normalized by electron tunneling length. \cite{maeda2005nanoscale}

The observed hysteresis behaviour with reversable parts of IV characteristic requires to add ‘’dry friction’’ term          $-\mu sign \frac{d\theta}{dt}$ resulting in a constant force equal $-\mu$ if the rode rotates with sign(x)=1 for $x > 0$ and sign(x)=-1 for $x < 0$.  Friction force $-\mu$ originates from multiple pinning on QDs and depends on their concentration (thus, ‘’friction’’ becomes a function of QDs concentration). For simplicity, we will consider only one bottleneck rod tilted by angle $\theta$ with respect to the shortest line connecting memristor terminals. 
Considering that the pinning electric potential tends to tilt the rod at angle $\theta_0$ while applied electric force/voltage stretches the rode towards zero angle (dipole interacting with constant electric field), we can write the phenomenological dynamical equation describing rod rotation:
\begin{align}
\tau \frac{d\theta}{dt} &= -\alpha \sin(\theta - \theta_0) - d_l V \sin(\theta) - \mu \text{sign}\left(\frac{d\theta}{dt}\right) \label{eq:first}\\
I &= \frac{V}{\exp(l(1-\cos\theta))} \label{eq:second}
\end{align}

Here we also assume the Ohm law to simulate I(V). The results of the simulations are shown in Figure 5 (a-d). Depending on the ratio between the nano-friction strength, \cite{saikia2018polyaniline} the repeating I-V loops could be ‘’one on top of the other’’ or ‘’one inside the other’’ in a qualitative agreement with experimental findings. Moreover, if after sweeping applied voltage from 0 to some values $V_1$, the voltage will decrease to value $V_2 < V_1$ and then increase to $V_3 > V_1$ we can reach any point inside of any I-V loop in full agreement with experimental observations. Such unprecedented ability to reach any point in I-V and the ability to keep the achieved values of resistance within certain voltage intervals due to rod nano-friction make the material to be very promising for designing continuously tenable memristive synapses needed for proper hardware implementation of deep learning circuits. \cite{yao2020fully,wang2017memristors}
\section{Conclusions}
In conclusion, we experimentally and theoretically studied the effect of concentration of GQDS in PANI matrix towards the memristive behavior in this nanocomposite. We have demonstrated that the NGQD-PANI nanocomposite exhibits a pinched non-volatile hysteresis.  The memristive behaviour of the nanocomposites have been explained by the influence of electric field that facilitates intramolecular charge transfer between NGQDs and PANI. This transfer induces partial delocalization of electrons within the $\pi$-conjugated system, enhancing or lowering the conductivity. However, not all trapped charges are fully released during the voltage cycles, contributing to diverse hysteresis behavior. Out findings stimulate further research towards understanding the memristive behavior in organic nanocomposites, which will lead to the development of various memristive systems that can be utilized for next-generation synaptic memory and weight units in deep neural networks.
\section{Acknowledgement}
This work was supported by the UK Engineering and Physical Sciences Research Council (EPSRC), grant no. EP/S032843/1.

\bibliography{Ref1}

\begin{figure}[h]
  \centering
  \includegraphics[width=2.0\linewidth]{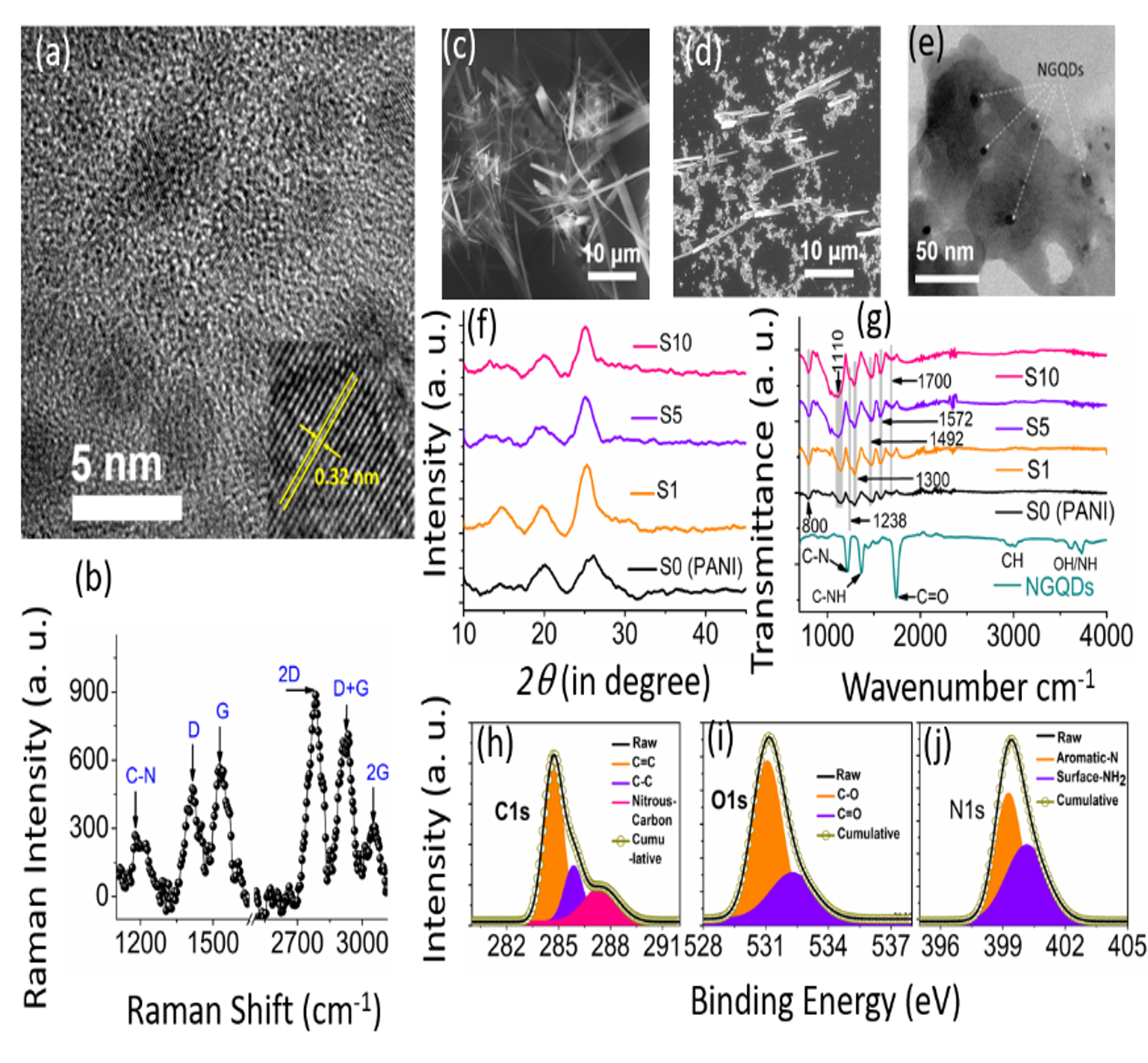}
  \captionsetup{width=2.1\linewidth}
  \caption{Structural features of the NGQD and NGQD-PANI nanocomposites: (a) bright field HR-TEM micrograph of the NGQDs. Inset: a high magnification image of a single dot with identifiable fringes. The Raman spectra of NGQD is shown in (b). FESEM images of bare PANI and 10$\%$ NGQD are shown in (c) and (d), respectively. (e) is the HRTEM image of the 10$\%$ NGQD-PANI nanocomposite showing the distribution of the QDs in PANI matrix. The XRD profile and the FTIR spectra of all the samples are shown in (f) and (g), respectively. Deconvoluted high-resolution XPS spectra corresponding of C 1s (h), O 1s (i) and N 1s (j), respectively.}
  \label{fig:fig1}
\end{figure}

\begin{figure}[h]
  \centering
  \includegraphics[width=1.8\linewidth]{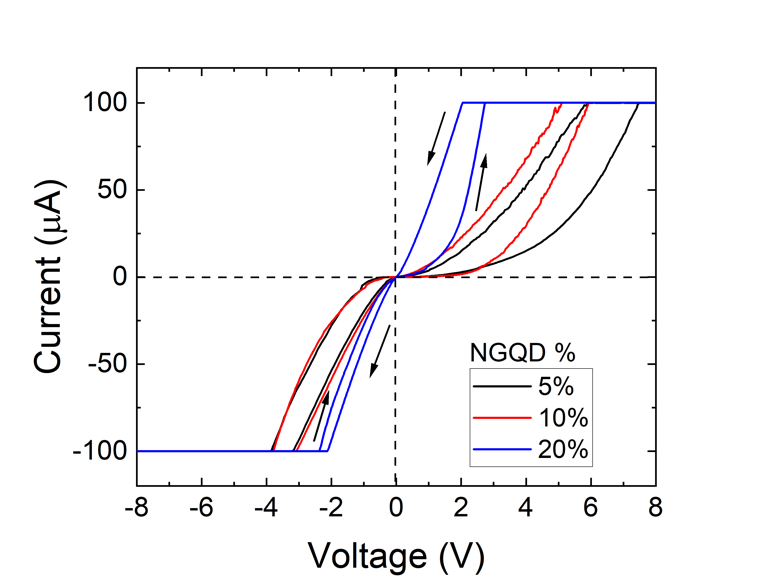}
  \captionsetup{width=1.8\linewidth}
  \caption{The room- temperature I-V characteristics of the pelletized NGQD-PANI composites for different NGQD concentration are shown in Figure 2 (a). These I-V characteristics are the average of 10 I-V cycles. All the I-V curves show a hysteresis behaviour. The increase in conductivity with NGQD concentration is associated with the increasing contribution of injected charge and quasi-alignment of the PANI matrix}
  \label{fig:fig2}
\end{figure}

\begin{figure}[h]
  \centering
  \includegraphics[width=1.8\linewidth]{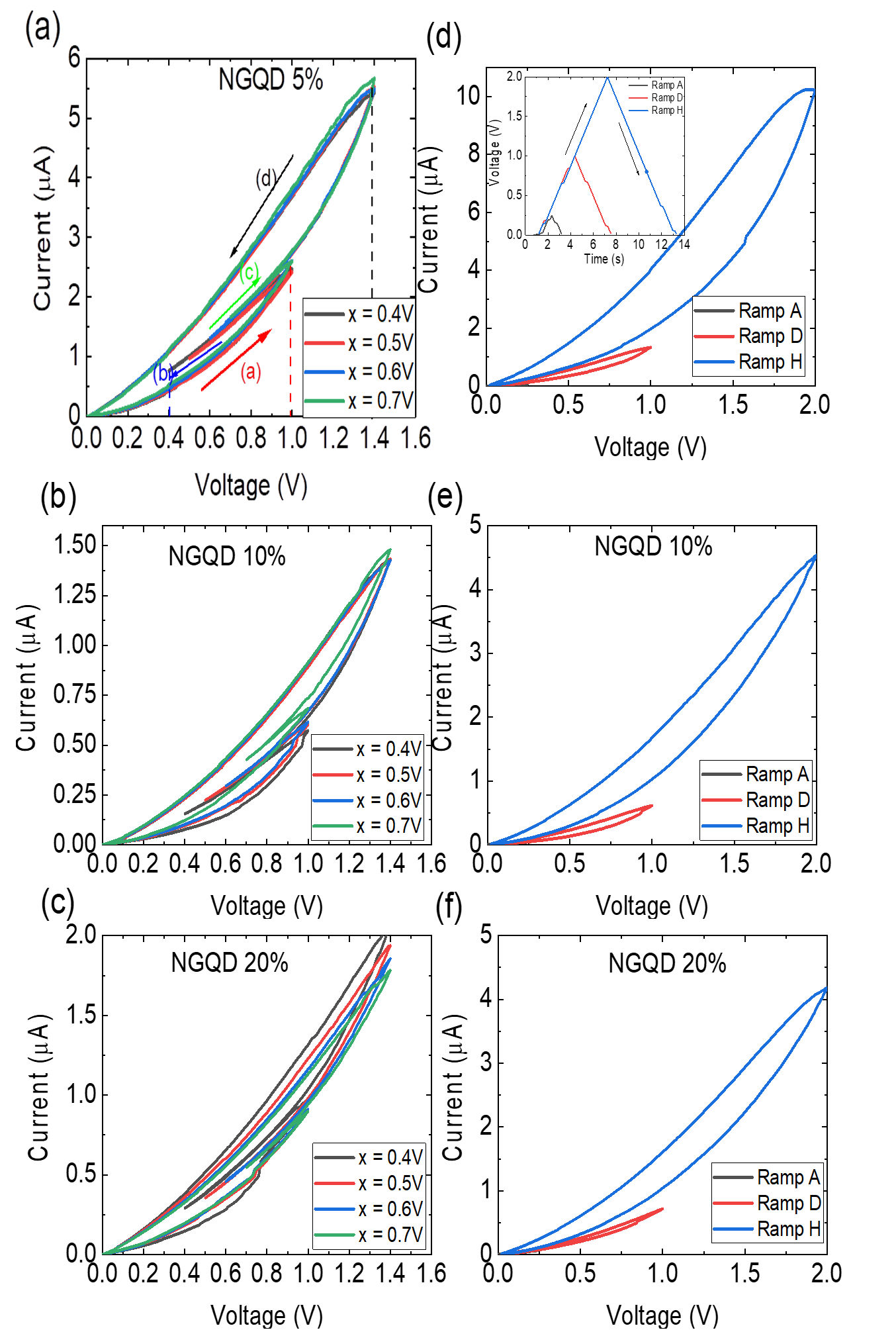}
  \captionsetup{width=1.8\linewidth}
  \caption{(a-c) RT IV loops for different NGQD concentration following this voltage cycle: 0V → 1V → x → 1.4V → 0V  (d-f) Triangular voltage from 0 to 2.5V in steps of 0.01V as shown in (d inset) was performed for all the samples. After the first triangular waveform the subsequent voltage waveforms were cycled from 0.2V to prevent a reset at 0V}
  \label{fig:fig3}
\end{figure}

\begin{figure}[h]
  \centering
  \includegraphics[width=1.8\linewidth]{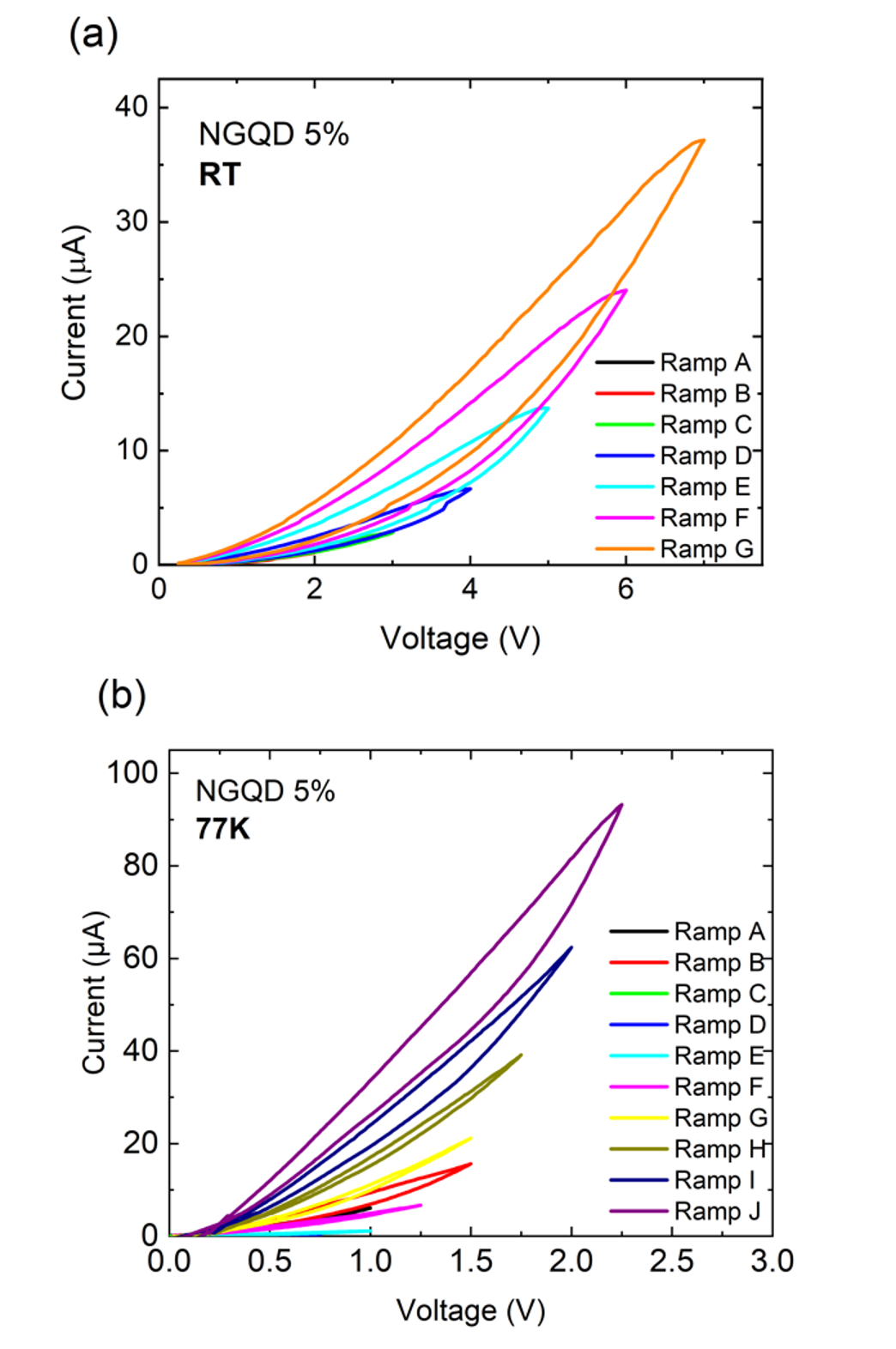}
  \captionsetup{width=1.8\linewidth}
  \caption{(a, b) I-V characteristics for RT and 77K respectively for 5$\%$ NGQD sample. The voltage was ramped as shown in the inset in Fig. 3 (d). }
  \label{fig:fig4}
\end{figure}

\begin{figure}[h]
  \centering
  \includegraphics[width=1.8\linewidth]{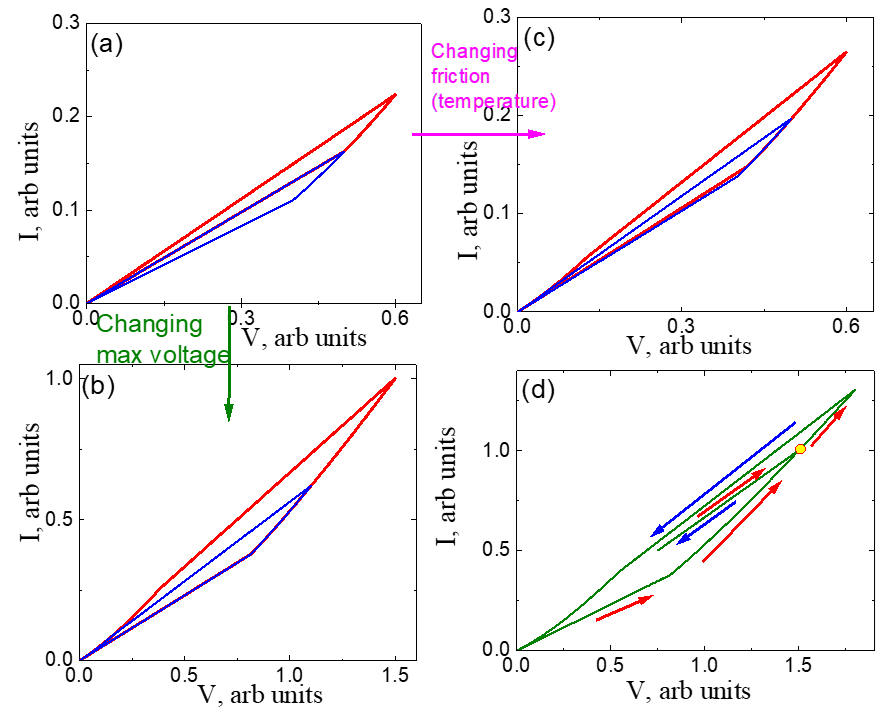}
  \captionsetup{width=1.8\linewidth}
  \caption{Current-voltage hysteresis clearly shows memristive behaviour. Depending on parameters of voltage sweep the successive loops (two voltage sweeps with amplitude of the second sweep slightly larger than the first sweep) can be ‘’one on top of the other’’ at low voltages (a), but this loop topology evolves towards ‘’one loop inside the other’’ with increasing sweep amplitudes (b) where we use friction coefficient to be $\frac{\mu}{\alpha} = 0.05$. Similar change of the topology occurs if we decrease friction$\frac{\mu}{\alpha} = 0.03$ (c) and keep all other parameters to be the same; this corresponds to increase of the temperature. Both observations are in perfect agreement with experimental finding. If for the case $\frac{\mu}{\alpha} = 0.05$, the voltage decreases and only, then, increase to maximum values, we will have the loop shown in (d), again in perfect agreement with experiment. For simulation we used inverse tunnelling parameter  l=100.  }
\label{fig:fig5}
\end{figure}
\end{document}